# A New Large-Well 1024x1024 Si:As Detector for the Mid-Infrared

A. K. Mainzer[1], John Hong[1], M. G. Stapelbroek[2], Henry Hogue[2], Dale Molyneux[2], Michael E. Ressler[1], Ernie Atkins[2], John Reekstin[2], Mike Werner[1], Erick Young[3]

[1]Jet Propulsion Laboratory, 4800 Oak Grove Dr., Pasadena, CA 91109; DRS Sensors & [2]Targeting Systems, 10600 Valley View Ave., Cypress, CA 90630; Steward Observatory, [3]University of Arizona, 933 N. Cherry Ave., Tucson, AZ, 85721

## ABSTRACT

We present a description of a new 1024x1024 Si:As array designed for ground-based use from 5 - 28 microns. With a maximum well depth of 5e6 electrons, this device brings large-format array technology to bear on ground-based mid-infrared programs, allowing entry to the megapixel realm previously only accessible to the near-IR. The multiplexer design features switchable gain, a 256x256 windowing mode for extremely bright sources, and it is two-edge buttable. The device is currently in its final design phase at DRS in Cypress, CA. We anticipate completion of the foundry run in the beginning of 2006. This new array will enable wide field, high angular resolution ground-based follow up of targets found by space-based missions such as the Spitzer Space Telescope and the Widefield Infrared Survey Explorer (WISE).

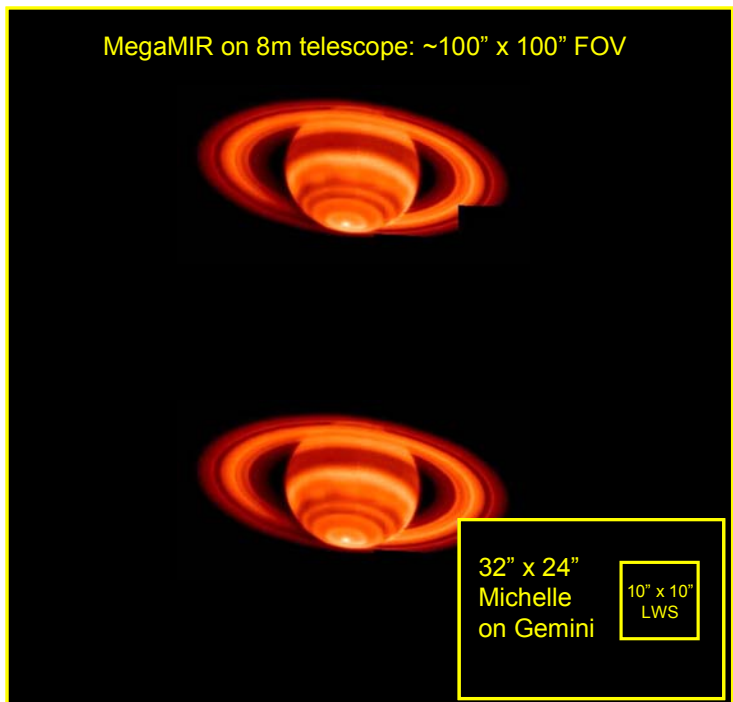

Figure 1: Saturn as seen by the Long Wavelength Spectrometer on the Keck Telescope at 17.65 μm. The largest yellow box represents the field of view possible if the MegaMIR array were incorporated into a camera on the Keck Telescope. The smaller boxes indicate the relative sizes of the Michelle and Long Wavelength Spectrometer (LWS) instruments (from Orton & Yanamandra-Fisher 2005). This image of Saturn took ~35 pointings with LWS; with the MegaMIR chip on an 8 m telescope, it would take only one. Not only can Saturn be imaged in one MegaMIR field of view, but it is possible to chop on-chip, doubling the effective integration time.

## 1. INTRODUCTION

The current state of the art mid-infrared focal plane arrays for high background applications have 128 x 128, 256 x 256, or a 320 x 240 pixel formats. Designed by DRS in Cypress, CA, our team is building the first 1024 x 1024 Si:As array optimized for ground-based use. This new chip, dubbed MegaMIR (Megapixel Mid-IR array), provides a 13-to-64x array format increase and benefits primarily from the recent developments in megapixel format, low background FPAs for space-based astronomy missions such as WISE (Wide Field Infrared Survey Explorer) and JWST (James Webb Space Telescope). The MegaMIR chip benefited from these designs, with key design modifications to accommodate the high background fluxes integral to ground-based use. With this new design, mid-infrared arrays now enter the megapixel realm formerly occupied only by near-infrared and visible devices.

### 1.1. SCIENTIFIC MOTIVATION

Although a ground-based mid-infrared camera cannot match the sensitivity of a cryogenic space-based telescope such as the Spitzer Space Telescope, there are many advantages in upgrading the largest telescopes in the world with what will be the largest mid-infrared array in the world. The aperture of the Spitzer Space Telescope is 0.85 m; hence, its spatial resolution is 2-7 arcsec, compared with the ~0.25 arcsec possible if the MegaMIR array were incorporated into a camera on an 8 m telescope.  This dramatic increase in spatial resolution enables a wide range of scientific topics to be addressed, including key follow up programs for both Spitzer and the upcoming WISE all-sky survey at 3.3, 4.7, 12, and 23 µm.  In addition, ground-based instruments featuring the MegaMIR array will serve as a bridge between the projected end of Spitzer's capabilities longward of 4.5 µm in 2009 and the launch of JWST.

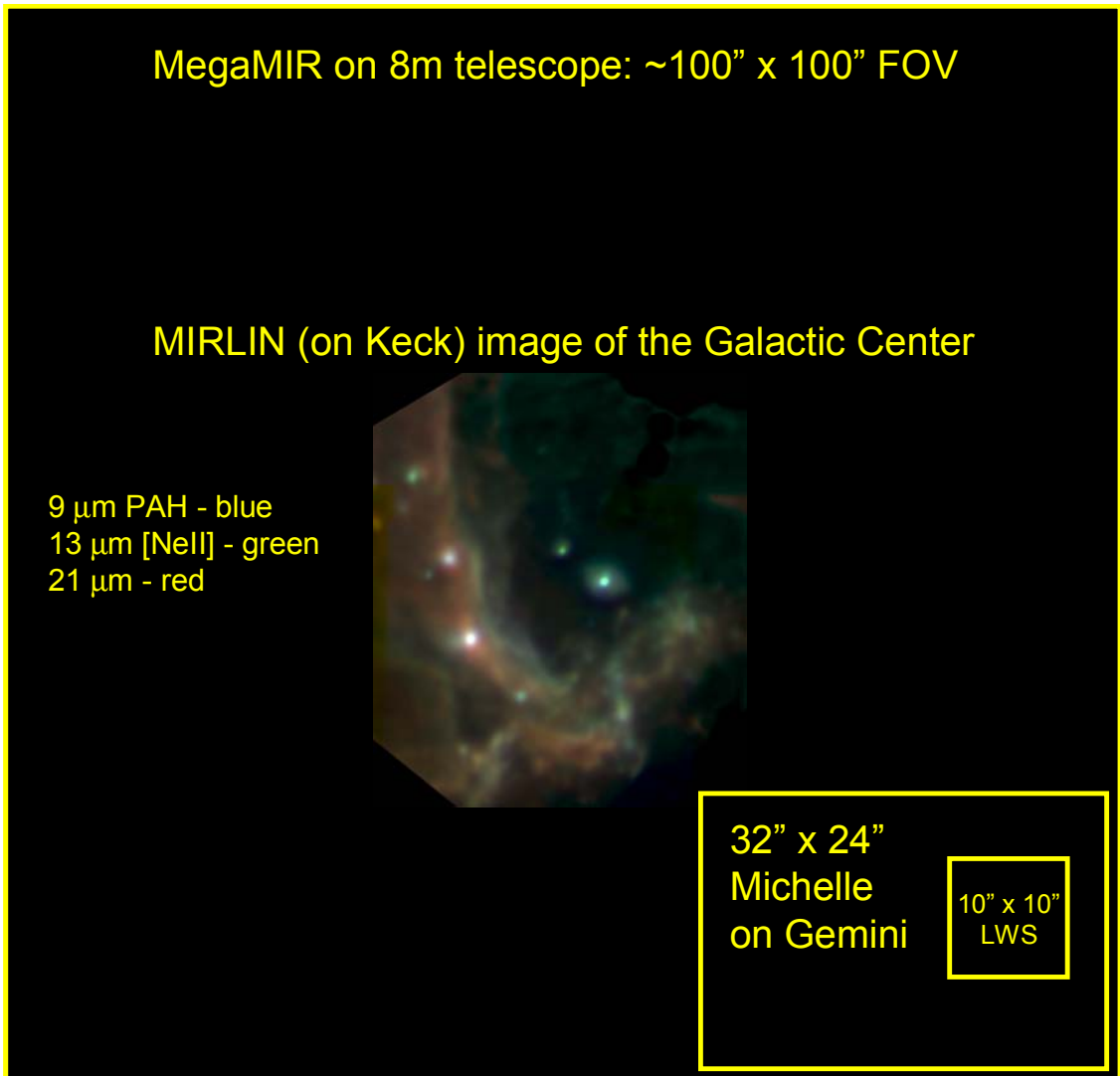

Figure 2: The Galactic Center observed with the MIRLIN camera on the Keck Telescope (Tanner et al. 2002, Ressler et al. 1994).  The field of view of the MegaMIR array on an 8 m telescope is shown for comparison, along with those of the Michelle and LWS instruments.

The MegaMIR 1024x1024 array will permit critical sampling over an ~100" field of view on an 8 m class telescope. This combination of high spatial resolution and a wide field will allow effective exploration of extended emission regions, including those taken from the Spitzer and WISE data bases.  In addition, whole disk images of the large planets - Jupiter, Saturn, Mars, and Venus - can be obtained in a single exposure.  For small extended sources, such as planetary nebulae, for example, the large field of view provided by the MegaMIR array will permit on-chip chopping and nodding for maximum efficiency.  For such investigations, the MegaMIR array will be one to two orders of magnitude more efficient than current MIR detectors which are no larger than 240 x 320 pixels.   For example, the

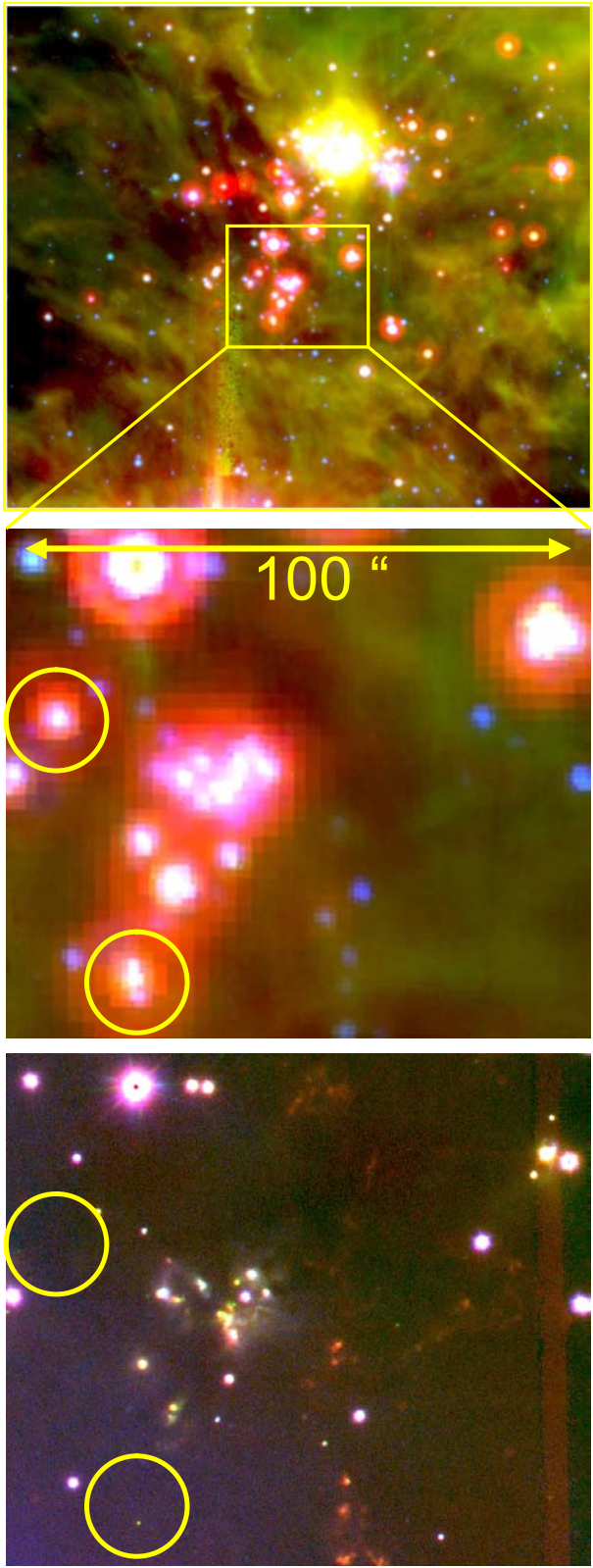

Figure 3a: Spitzer image of a portion of a nearby galactic star formation complex. The color code is blue=3.6 µm, green=8.0 µm, and red=24 µm. 3b) Blow up of a particularly interesting and crowded region of the Spitzer image. The field of view is 100 x 100 arcsec, equal to that of MegaMIR. 3c) A ground-based H (blue), K (green), and molecular hydrogen (red) of the same region at 0.35 arcsec resolution. The circles indicate objects that are visible in the mid-infrared, but not in the near-IR, illustrating the need for dust penetrating mid-IR observations.

Michelle camera in use at the Gemini observatory has only a 24x32" field of view. Figure 1 depicts Saturn as seen in one field of view of the MegaMIR array on an 8 m telescope compared to the Michelle instrument on Gemini and the Long Wavelength Spectrometer (LWS) on Keck. This image of Saturn (Orton & Yanamandra-Fisher 2005) was produced using the 128 x 128 Si:As array in LWS, taking ~35 pointings. A single-frame image of Saturn would facilitate time-variability studies over the rotational period of the planet. Figure 2 illustrates the improvements to imaging of the Galactic Center that the MegaMIR chip on an 8 m telescope could provide.

Figure 3a shows a Spitzer image of a portion of a nearby star formation complex. The color code is blue=3.6 µm, green=8.0 µm, and red=24 µm. Figure 3b shows a blow up of a particularly interesting and crowded region of the Spitzer image. The field of view is 100 x 100 arcsec, equal to that of MegaMIR on an 8 m telescope. Figure 3c shows a ground-based H (blue), K (green), and molecular hydrogen (red) image of the same region at 0.35 arcsec resolution, showing a complex cluster of sources with a dynamical lifetime of ~20,000 yrs. MegaMIR can measure the thermal infrared emission from these objects at similar spatial resolution, determining the luminosity and dust mass of each of these protostars. As a flux reference, the bright source on the left edge of Figure 3b that is not visible in 3c is roughly 85 mJy at 8 µm and thus easily detectable with MEGAMIR. At the distance of ~300 pc, this corresponds to less than a solar luminosity in the 10 µm band.

Extragalactic science also stands to benefit from the MegaMIR array. Figure 4 depicts a Spitzer Space Telescope image of the nearby spiral galaxy M81 (Gordon et al. 2004). As shown in the figure, the MegaMIR array with its wide field of view on an 8 m telescope will enable rapid follow up of multiple extragalactic star forming regions with 10x higher spatial resolution than Spitzer can provide. The MegaMIR array is sensitive enough to see individual pre-main sequence stars and evolved dusty AGB stars at a distance of 1 Mpc at 10 µm (e.g., the distance of M31). MegaMIR's angular resolution is high enough to allow detailed studies of extended star forming regions discovered by Spitzer.

Grism spectroscopy using the MegaMIR array will enable important investigations with in planetary, galactic, and extragalactic science. Since objects that are confused in the Spitzer Infrared Spectrograph's 3 arcsec beam can now be sorted out into their individual components, e.g. galactic nuclei/disks, the polar regions and individual cloud complexes in Jupiter and Saturn's atmospheres, etc. Interestingly, MegaMIR's resolution of Jupiter on an 8 m telescope (900 km) exceeds that of the Voyager IRIS instrument, the Galileo PPR, and the Cassini CIRS – all of which were actually at the planet. Low resolution (R~500) spectroscopy of Jupiter with high spatial resolution over a large field of view will enable studies of cold airmass polar vortices, spectroscopic searches for gasses, polar auroras, the para-to-ortho $H_2$ fraction, and time-varying studies of storms and merging giant vortices. Figure 6 depicts a sample of low resolution spectroscopy using the MegaMIR chip compared with the Spitzer low resolution slit. The Spitzer Infrared Spectrometer (IRS) low resolution module has a 3.7" wide slit, compared with the 0.3" slit possible with MegaMIR.

The results of our group's work with the MIRLIN camera (Ressler et al. 1994) together with those obtained by other groups dramatize the scientific power of a thermal infrared camera on a large telescope.

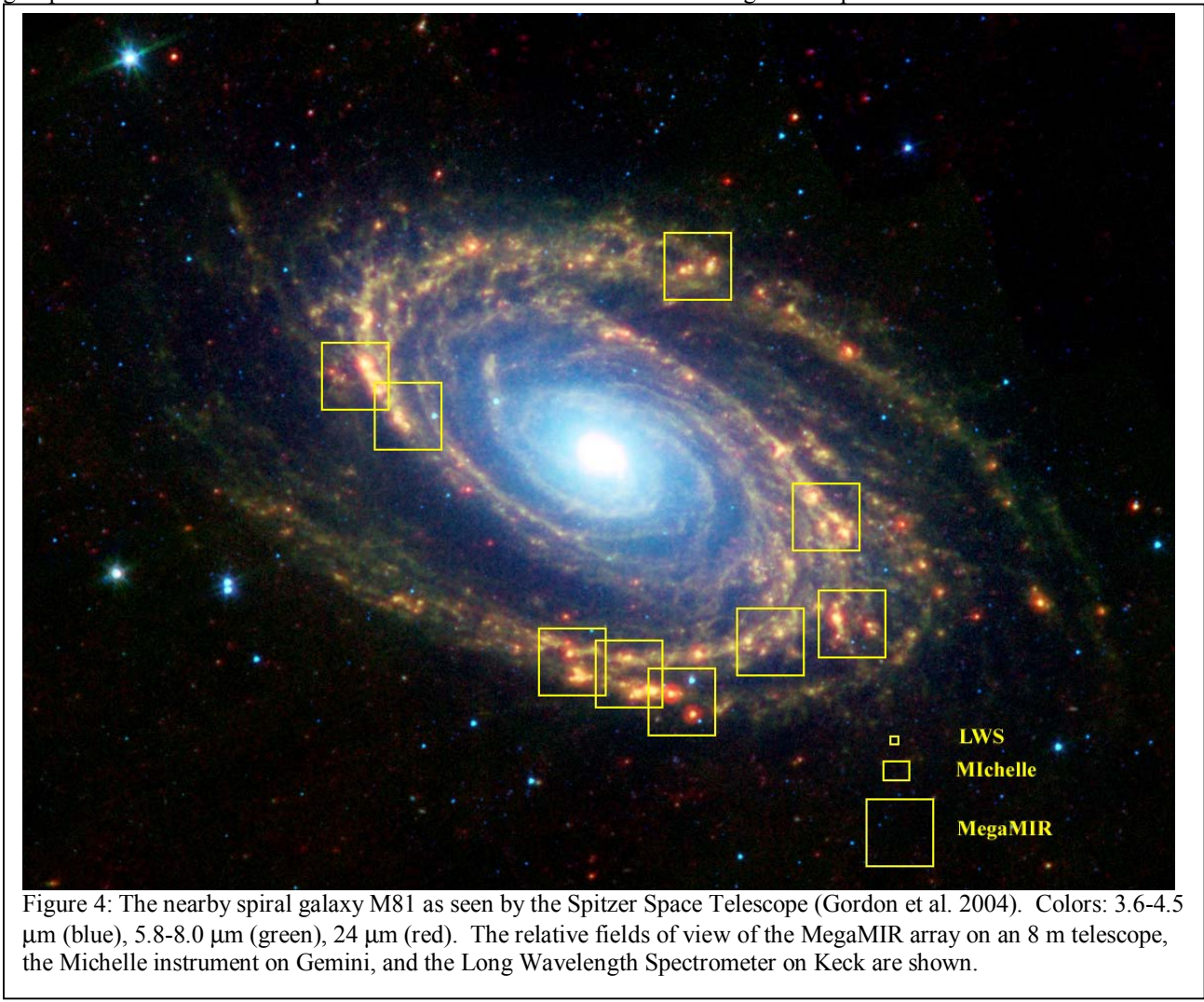

Figure 4: The nearby spiral galaxy M81 as seen by the Spitzer Space Telescope (Gordon et al. 2004). Colors: 3.6-4.5 µm (blue), 5.8-8.0 µm (green), 24 µm (red). The relative fields of view of the MegaMIR array on an 8 m telescope, the Michelle instrument on Gemini, and the Long Wavelength Spectrometer on Keck are shown.

## 2. MegaMIR FPA

The MegaMIR focal plane array (FPA) is a Si:As blocked impurity band (BIB) detector array hybridized with indium bump bonds onto a CMOS multiplexer. The multiplexer features four independent 512 x 512 quadrants using direct injection unit cells. Four outputs are provided per quadrant for a total of 16 outputs. A non-destructive read mode is possible. The array is two-edge buttable, allowing it to be dense-packed (see Figure 7) to form a 2048 x 2048 array. The array can also be operated in a 256 x 256 high speed window mode for extremely bright sources. Here, each quadrant can be windowed to a 128 x 128 corner to produce a 256 x 256 centered window for the four quadrants packed together. The integration capacitance can be switched providing for two very different maximum well depths, 5e6 $e^-$ and 1e5 $e^-$. The large well mode is suitable for broad and narrow band imaging, and the low gain mode is intended for spectroscopy.

Table 1 gives the preliminary MegaMIR array specifications for the large well mode.

| Parameter | Value | Units | Comments |
|---|---|---|---|
| Format | 1024 x 1024 | | 2 edge buttable to form 2048x2048 |
| Integration Mode | Ripple | | |
| Pixel Pitch | 18 | µm | |
| Detector Material | Si:As | | |
| Band | 5-28 | µm | |
| Integration Control | Variable | | |
| Read Noise | < 1000 | $e^-$ | 10 msec integration time |
| Dark Current | < 12,800<br>< 10 | $e^-$/s<br>e-/s | at 10K<br>at 6K |
| QE | > 57% | | |
| Well Capacity | ≤5E+06 Max<br>≥1E+05 Min | $e^-$ | Controlled by gain select |
| Operability | > 99% | | |
| Non-uniformity | < 2% | | Limited by measurement accuracy |
| Linearity | < 10% | | |
| Frame Rate | < 100 | Frames/sec | |
| Number of Outputs | 16 | | |
| Data Rate | ≤ 7E6 | Pix/s/output | |

Table 1: MegaMIR array specifications

## 3. BLOCKED IMPURITY BAND DETECTORS

Because BIB detector physics, models, and performance have been previously documented (Petroff et al. 1984, Petroff et al. 1985, Reynolds et al. 1989, Stapelbroek et al. 1984), only the salient features of these devices are summarized here. BIB detectors effectively use the hopping conductivity phenomenon associated with "impurity banding" in relatively-heavily-doped semiconductors (specifically single-crystalline Si:As or Si:Sb for this discussion). In particular, a BIB detector comprises two thin epitaxial layers placed between planar electrical contacts; a relatively heavily-doped infrared-active layer and an undoped or lightly-doped blocking layer. Figure 8 illustrates the structure of a BIB detector constructed to allow back-illumination through a transparent silicon substrate. All active detector structures are fabricated on the front side of the undoped single-crystal substrate wafer.

Both low- and high-flux versions of the BIB detector have been refined over a number of years. Significant adjustments to the detector design have to be made to optimize for high-flux operation. These adjustments included changes in doping profiles and layer thicknesses and tailoring buried contact resistivity to allow for the larger current densities experienced in the higher photon flux environment.

Additional detector design and process improvements were also made to eliminate excess low-frequency noise (ELFN) that has been observed in BIB detectors operated under high infrared background conditions (Stapelbroek et al. 1984). This noise originates from space-charge fluctuations in the detector's blocking layer and manifests itself as a "white" noise at low frequencies with a roll-off at higher frequencies, as illustrated by the curve labeled "EARLY HFPAs" in Figure 9.  The noise at higher frequencies is shot noise due to the photon background to which the noise curves are normalized. The improvement realized in high-background detectors developed for a passive seeker application is evident in the curve labeled "IMPROVED HFPAs," which has essentially no ELFN. The data were obtained on fully-functional HFPAs. The roll-off frequency for ELFN is directly proportional to the background flux; thus, ELFN is generally not an issue for low-background arrays.

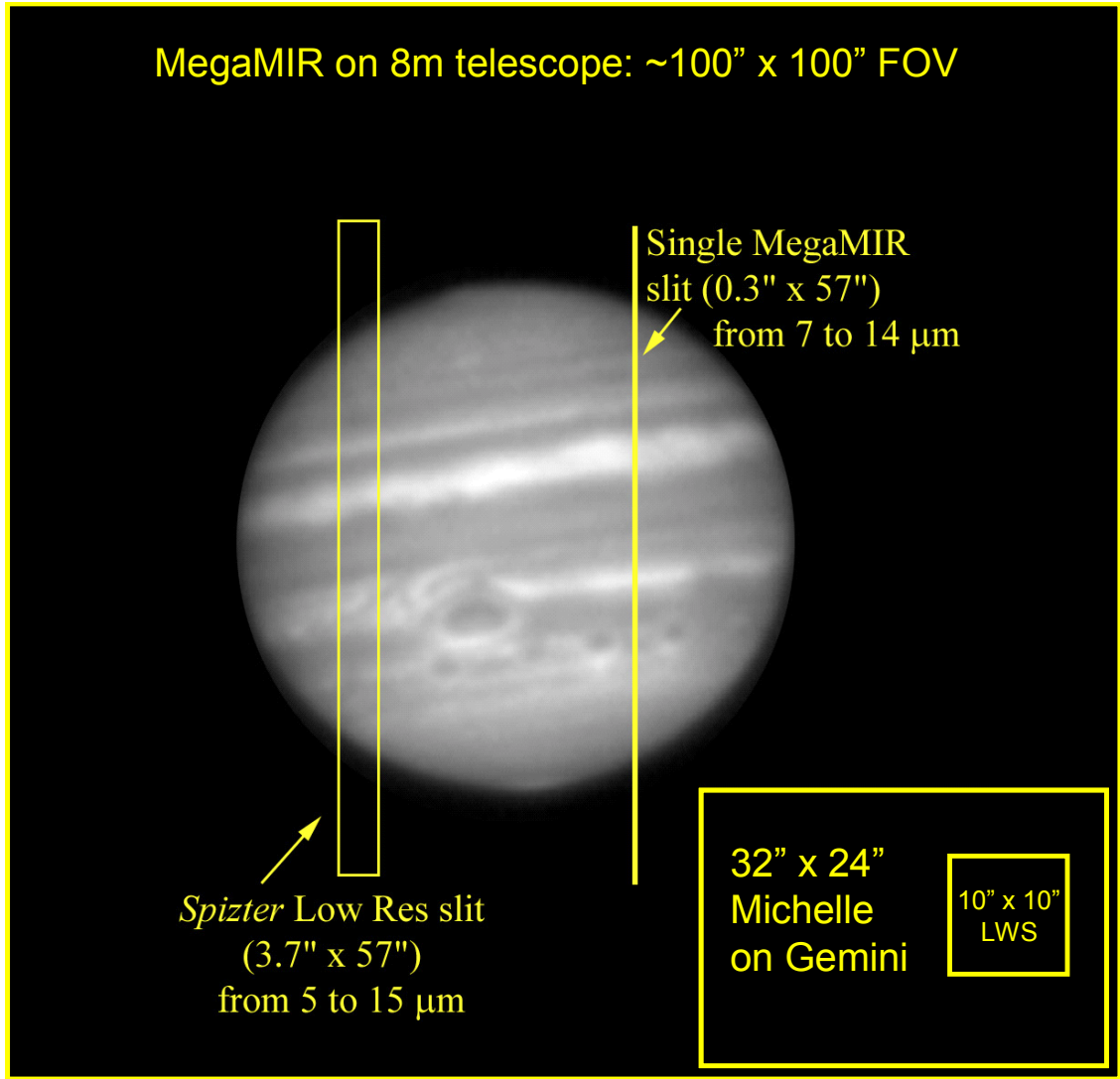

Figure 6: Jupiter at 10.3 µm as seen by NASA's Infrared Telescope Facility (Fisher et al. 1999). Using the MegaMIR array in a low resolution spectrometer on an 8 m class telescope will result in 10x better spatial resolution than the Spitzer Space Telescope's IRS over the entire disk of Jupiter.

## 4. ACKNOWLEDGEMENTS

This work was funded by the Jet Propulsion Laboratory, operated for the National Aeronautics and Space Administration by the California Institute of Technology.

## 5. REFERENCES

1. G. S. Orton, P. A. Yanamandra-Fisher, "Saturn's Temperature Field from High-Resolution Middle-Infrared Imaging," Science 307, 696 (2005).
2. M. E. Ressler, M. W. Werner, J. Van Cleve, and H.~A. Chou, "The JPL deep-well mid-infrared array camera," Experimental Astronomy 3, 277 (1994).
3. M. D. Petroff and M. G. Stapelbroek, "Responsivity and Noise Models of Blocked Impurity Band Detectors," Proc. IRIS Specialty Group on Infrared Detectors (August 1984, Seattle, WA).
4. M. D. Petroff and M. G. Stapelbroek, "Spectral Response, Gain, and Noise Models for IBC Detectors," Proc. IRIS Specialty Group on Infrared Detectors ( August 1985, Boulder, CO).
5. D. B. Reynolds, D. H. Seib, S. B. Stetson, T. Herter, N. Rowlands, and J. Schoenwald, "Blocked Impurity Band Hybrid Infrared Focal Plane Arrays for Astronomy," IEEE Trans. Nucl. Science 36, 857 (1989).
6. M. G. Stapelbroek, M. D. Petroff, J. J. Speer, and R. Bharat, "Origin of Excess Low Frequency Noise at Intermediate IR Backgrounds in BIB Detectors," Proc. IRIS Specialty Group on Infrared Detectors (August 1984, Seattle, WA).
7. Tanner, A.; Ghez, A. M.; Morris, M.; Becklin, E. E.; Cotera, A.; Ressler, M.; Werner, M.; Wizinowich, P. "Spatially Resolved Observations of the Galactic Center Source IRS 21" 2002 ApJ 575, 860
8. Gordon, K. D.; Pérez-González, P. G.; Misselt, K. A.; Murphy, E. J.; Bendo, G. J.; Walter, F.; Thornley, M. D.; Kennicutt, R. C., Jr.; Rieke, G. H.; Engelbracht, C. W.; Smith, J.-D. T.; Alonso-Herrero, A.; Appleton, P. N.; Calzetti, D.; Dale, D. A.; Draine, B. T.; Frayer, D. T.; Helou, G.; Hinz, J. L.; Hines, D. C.; Kelly, D. M.; Morrison, J. E.; Muzerolle, J.; Regan, M. W.; Stansberry, J. A.; Stolovy, S. R.; Storrie-Lombardi, L. J.; Su, K. Y. L.; Young, E. T. "Spatially Resolved Ultraviolet, H , Infrared, and Radio Star Formation in M81" 2004 ApJS 154, 215
9. B. M. Fisher, G. S. Orton, P. Yanamandra-Fisher, M. Ressler, P. Fukumura-Sawata, W. Golisch, D. Griep, J. Spencer, L. Sromovsky, F. Fry "Metastable cyclones in Jupiter: Interactions between white ovals and the rise of very dark spots in 1998 and 1999" 1999 Amer. Astron. Soc. 31, 1158

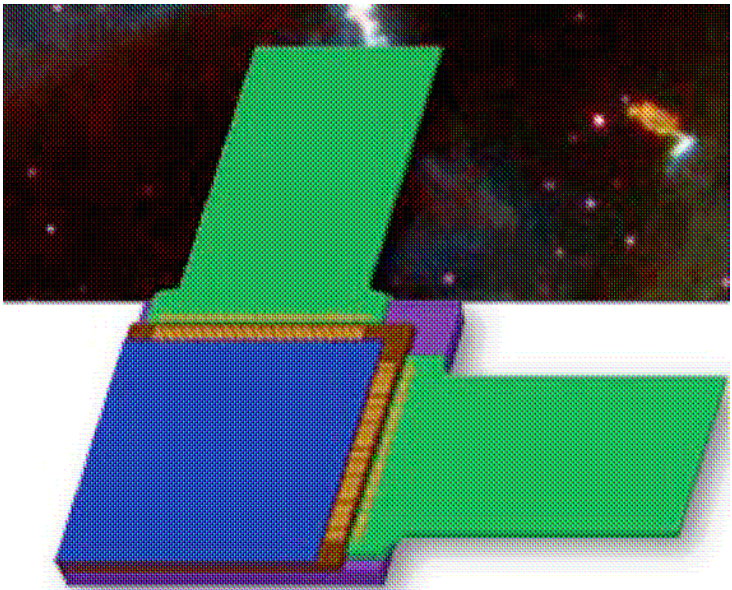

Figure 7: The MegaMIR 1024 x 1024 array, with pinout arranged to allow four arrays to be butted together to form a 2048 x 2048 array.

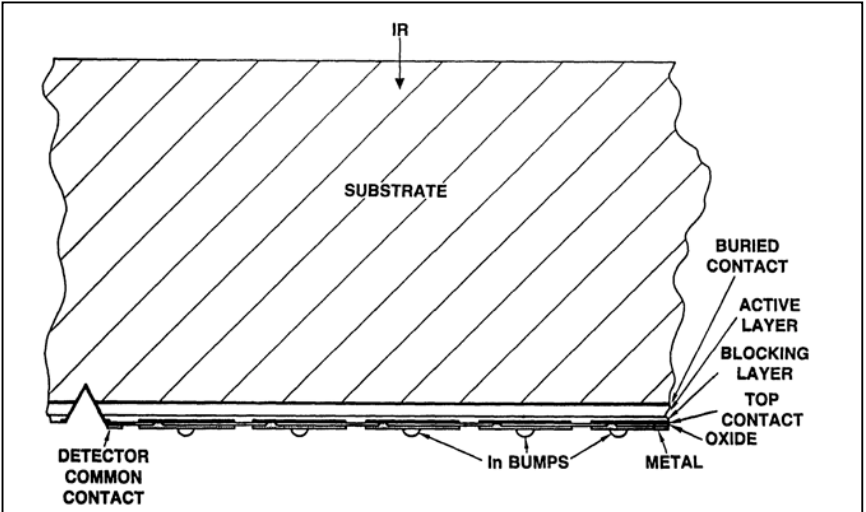

Figure 8. Schematic cross section of a back-illuminated BIB detector array. The drawing is approximately to scale for typical pixel sizes.

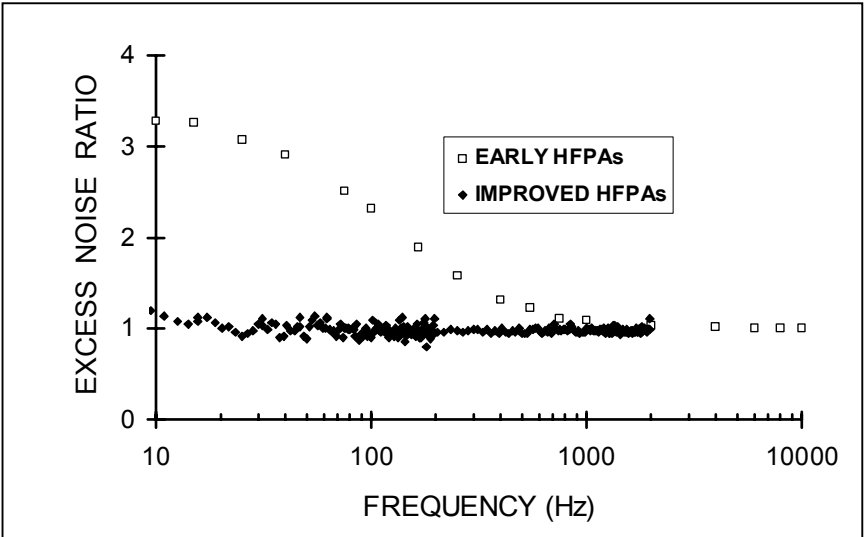

Figure 9. Ratio of HFPA noise to photon shot noise as a function of frequency. Excess low-frequency noise has been eliminated in the improved HFPAs.